\documentclass[fleqn]{2023SCGE}
\usepackage{amsmath}
\usepackage{aas_macros}
\usepackage{graphicx}
\usepackage{seqsplit}
\usepackage{natbib}

\newcommand{\gray}{$\gamma$-ray\ }

\newcommand{\grays}{$\gamma$-rays\ }

\newcommand{\signi}[1]{$#1\nobreak\,\sigma$}

\begin{document}

\ensubject{subject}

\ArticleType{Article}
\SpecialTopic{SPECIAL TOPIC: }
\Year{2024}
\Month{XXXX}
\Vol{XX}
\No{XX}
\DOI{??}
\ArtNo{XXX}
\ReceiveDate{ }
\AcceptDate{ }

\title{Study of Ultra-High-Energy Gamma-Ray Source 1LHAASO J0056+6346u and Its Possible Origins}

\author{LHAASO Collaboration\footnote{Corresponding author: G.W. Wang (wangguangwei@mail.ustc.edu.cn), B. Liu (liubing@pmo.ac.cn), W.Y. Cao (caowy@mail.ustc.edu.cn), Y.H. Yu (yuyh@ustc.edu.cn)} \\(The LHAASO Collaboration authors and affiliations are listed after the references.)}{}%

\AuthorMark{Zhen Cao}
\AuthorCitation{Zhen Cao, et al}






\abstract{
We report a dedicated study of the newly discovered extended UHE \gray source 1LHAASO J0056+6346u. Analyzing 979 days of LHAASO-WCDA data and 1389 days of LHAASO-KM2A data, we observed a significant excess of \gray events with both WCDA and KM2A. Assuming a point power-law source with a fixed spectral index, the significance maps reveal excesses of \signi{12.65}, \signi{22.18}, and \signi{10.24} in the energy ranges of 1–25 TeV, 25–100 TeV, and $>$ 100 TeV, respectively. We use a 3D likelihood algorithm to derive the morphological and spectral parameters, and the source is detected with significances of \signi{13.72} by WCDA and \signi{25.27} by KM2A. The best-fit positions derived from  WCDA and KM2A data are {(R.A. = $13.96^\circ\pm0.09^\circ$, Decl. = $63.92^\circ\pm0.05^\circ$)} and  {(R.A. =  $14.00^\circ\pm0.05^\circ$, Decl. = $63.79^\circ\pm0.02^\circ$)}, respectively. The angular size ($r_{39}$) of 1LHAASO J0056+6346u is {$0.34^\circ\pm0.04^\circ$} at 1-25 TeV and {$0.24^\circ\pm0.02^\circ$} at $\textgreater$ 25 TeV. The differential flux of this UHE \gray source can be described by an exponential cutoff power-law function: $(2.67\pm0.25) \times 10^{-15}\,(E\, /\, {20 {\rm\ TeV}})^{(-1.97\pm0.10)}\,{e}^{-E\,/\,{(55.1\pm7.2) \rm\ TeV}} \rm\  TeV^{-1} cm^{-2} s^{-1}$. 
To explore potential sources of \gray emission, we investigated the gas distribution around 1LHAASO J0056+6346u.  1LHAASO J0056+6346u is likely to be a TeV PWN powered by an unknown pulsar, which would naturally explain both its spatial and spectral properties.  
Another explanation is that  this UHE \gray source might be associated with gas content illuminated by a nearby CR accelerator, possibly the SNR candidate G124.0+1.4.
}



\maketitle


\begin{multicols}{2}

\section{Introduction}\label{sec:intro}

Recently, the LHAASO experiment opened the window of Ultra High Energy (UHE, $E > 100{\,\rm TeV}$) \gray astronomy with the discovery of 43 sources above 100 TeV   \citep{LHAASOJ0341,caoUltrahighenergyPhotonsPetaelectronvolts2021a,lhaaso_cat1}.
The horizon of UHE astronomy is well defined because of the strong absorption of \gray photons above 100 TeV, reducing the mean free path of these photons to several hundred kiloparsecs. Consequently, all these sources detected by LHAASO have little doubt of Galactic origin. However, the classification of these sources is highly diverse. The possible associations are primarily Supernova Remnants (SNRs),  Pulsar Wind Nebulae (PWNe), stellar clusters, among others.

 \begin{figure*}[ht!]
     \centering
     \includegraphics[width=\textwidth]{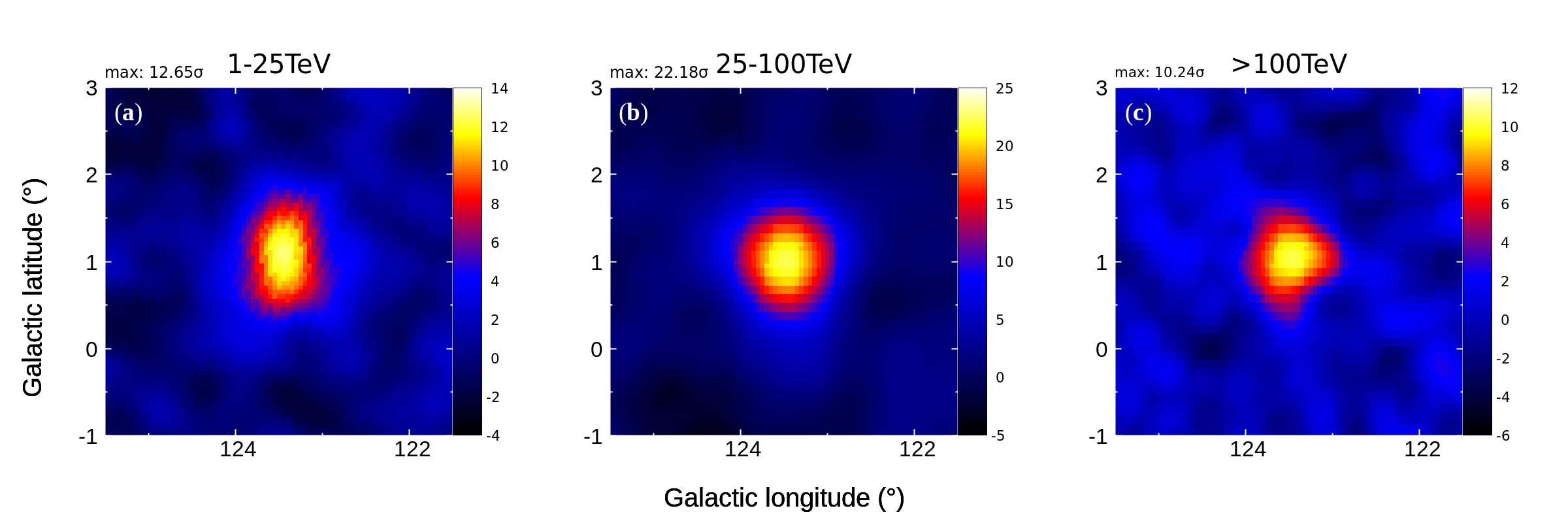}
     \caption{ Significance maps of 1LHAASO J0056+6346u (zoomed-in view smaller than ROI) derived from the prior model, with GDE subtracted. These figures show the significance maps for the energy ranges (a) 1 -- 25 TeV\@, (b) 25 -- 100 TeV\@, and (c) above 100 TeV, respectively.}
     \label{fig:km2a significance map}
 \end{figure*}

With the accumulated exposure, LHAASO has detected more sources in the Galactic plane and published its first catalog of \gray sources  \citep{lhaaso_cat1}, some of which remain unidentified and require further investigation.  In this paper, we present a dedicated study of the unidentified UHE \gray source 1LHAASO J0056+6346u. This source shows no spatial coincidence with any known SNRs or PSRs along the line of sight.  To investigate its origin, we analyzed data from both the WCDA and KM2A detectors of LHAASO (accumulated over a longer period up to January 2024) and performed a detailed multiwavelength analysis of the source surroundings. The paper is structured as follows. In Sec.~2, we present the results of the analysis for the WCDA and KM2A data. In Sec.~3, we describe the multiwavelength observations, especially the \textit{Fermi}-LAT GeV observations and the gas distributions. In Sec.~4, we discuss the possible origin of this source. Finally, we summarize this work in Sec.~5.

\section{LHAASO Data Analysis and Results} \label{sec:LDAAR}

The  \gray data below $25\rm\ TeV$ were collected by the whole WCDA, covering an area of 78,000 m$^2$, from March 2021 to January 2024 \citep{aharonianPerformanceLHAASOWCDAObservation2021}. The \gray data above $25\rm\ TeV$ were collected from different phases of the KM2A array: half the array since December 2019, three-quarters of the array from December 2020 to July 2021, and the full KM2A array from July 2021 to {January} 2024. After the data quality check   \citep{quality_check}, the total effective observation time is {979} days for WCDA and {1389} days for KM2A. We used the same event reconstruction and selection criteria as in  \citep{crabcpc, aharonianPerformanceLHAASOWCDAObservation2021}. 

The KM2A data sets are divided into five groups per logarithmic energy decade based on the reconstructed energy. WCDA uses the number of triggered PMT units ($N_{hits}$) as the shower energy estimator. All the events are grouped into six bins of $N_{hits}$, namely $\left[60, 100\right)$, $\left[100, 200\right)$, $\left[200, 300\right)$, $\left[300, 500\right)$, $\left[500, 800\right)$, and $\left[800, 2000\right]$. For each data group,  the sky map in celestial coordinates (right ascension and declination) is divided into a grid of $0.1^\circ\times0.1^\circ$ each pixel filled with the number of detected events according to their reconstructed arrival direction. The number of cosmic ray (CR) background events in each pixel is estimated using the “direct integration method'' \citep{Fleysher_2004},  which estimates the background by analyzing events with consistent directional alignment in local coordinates but differing arrival times. In this work, 10 hours of data are integrated to estimate detector acceptance for different directions. This integrated acceptance, combined with the event rate, is used to calculate background event counts in each pixel, resulting in a background map.

\begin{figure*}[ht!]
    \centering
    \includegraphics[width=\textwidth]{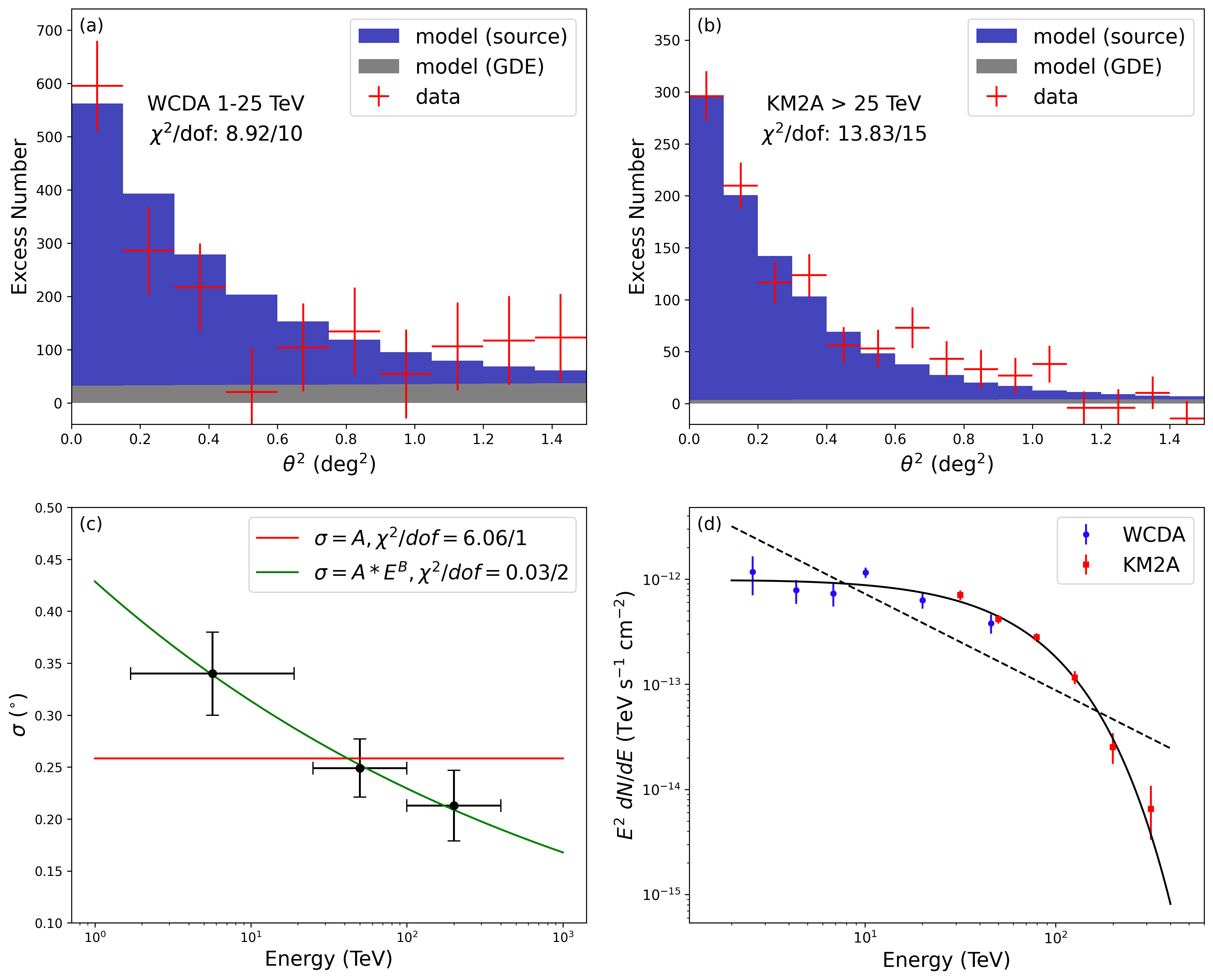}
    \caption{Extension and spectrum of 1LHAASO J0056+6346u. (a-b) Distribution of events as a function of square of the angular offset to 1LHAASO J0056+6346u for data (red points, the observed number of events minus the CR background), expected number of events from the source (blue region),  as well as the GDE (grey region). Panel (a) displays the events detected by WCDA in the $1-25\,{\rm TeV}$ range, while panel (b) shows the events detected by KM2A with energy above $25\,{\rm TeV}$. (c) Energy dependence of source extension, with the red line representing the energy-independent model and the green line representing the power-law model.
    (d) The spectrum of 1LHAASO J0056+6346u, with the dotted line representing the single power-law fit and the solid line representing the ECPL fit.}
    \label{fig:km2a angular distribution}
 \end{figure*}

In this work, we use the Test Statistic (TS) to evaluate the goodness of fit for different models. TS is defined as 
\begin{equation}
 {\rm TS} = \frac{\mathcal{L}_{1}}{\mathcal{L}_{0}} = \frac{\displaystyle \max_{\bm{\theta}_1} \prod_{i}^{}{\rm Poisson}(N_i^{\rm obs}, N_i^{\rm exp1})}{\displaystyle \max_{\bm{\theta}_0}\prod_{i}^{}{\rm Poisson}(N_i^{\rm obs}, N_i^{\rm exp0})} 
\tag{1},
\end{equation}
where $\mathcal{L}_{1}$ is the maximum likelihood value for the alternative hypothesis (the hypothesis under test), while ${\mathcal{L}_0}$ is the maximum likelihood value of the null hypothesis. $\bm{\theta}_1$ and $\bm{\theta}_0$ are the free parameters in the alternative hypothesis and the null hypothesis, respectively. `i' is the index of each bin with specific energy and position. $N_i^{\rm obs}$ is the observed number of events in each bin. $N_i^{\rm exp1}$ and $N_i^{\rm exp0}$ are the expected number of events in each bin of the alternative hypothesis and the null hypothesis, respectively. The expected number of events accounts for contributions from the estimated CR background, the Galactic Diffuse Emission (GDE), and individual sources in each model. The $\mathrm{Poisson}$ term quantifies the statistical probability of  $N_i^{\rm obs}$, given $N_i^{\rm exp}$.  According to Wilks' Theorem  \citep{wilks}, the TS value follows a chi-square distribution with degrees of freedom equal to the difference in the number of free parameters between the null and alternative hypotheses.

To identify potential new sources, we create TS maps. The alternative hypothesis assumes the presence of an additional point source at each tested position in addition to the CR background and the GDE, while the null hypothesis includes only the CR background and the GDE. The additional point source is modeled with a power-law spectrum with a fixed spectral index of $3.0$ for KM2A and $2.6$ for WCDA, consistent with the first LHAASO catalog \citep{lhaaso_cat1}. Flux is the only free parameter. According to Wilks' Theorem,  $\pm\sqrt{\rm TS}$ represents the significance of the alternative hypothesis. The significance maps are shown in Fig.~\ref{fig:km2a significance map}, in which the GDE uses a fixed spectrum, consistent with parameters obtained from the prior model. The highest significance values in the maps are \signi{12.65} in the energy band of 1--25 TeV, \signi{22.18} in the energy band of 25--100 TeV, and \signi{10.24} in the energy band of $>$ 100 TeV, respectively.

We use a 3D likelihood algorithm to simultaneously determine the morphology and spectrum of the source. The alternative hypothesis includes the individual source, the GDE, and the CR background, while the null hypothesis includes only the CR background. In our model, the source is assumed to have a two-dimensional Gaussian morphology with variable position and extension, and a power-law spectrum with variable flux and spectral index. For KM2A, we also test an Exponential Cutoff Power-Law (ECPL) spectrum which includes an additional cutoff parameter. To estimate the impact of GDE on our analysis, we incorporate a spatial template based on the gas distribution.  Similar to the approach in \citet{lhaaso_cat1}, we assumed the diffuse CRs are uniform in the ROI, and the GDE, produced by diffuse CRs, is proportional to the gas column density map which is derived from the PLANCK dust opacity map \citep{planck_dust_2014, planck_dust_2016}. The spectrum of the GDE template was left free in the likelihood analysis. The region of interest (ROI) for this likelihood fit is defined as a circular disk with a radius of 6$^\circ$, centered at (R.A. = 14.0$^\circ$, Decl. = 63.5$^\circ$). We determined the parameters that yield the maximum TS value.

KM2A observations revealed an extended source with an intrinsic extension ($r_{39}$) of $0.24^\circ\pm0.02^\circ$ at the best-fit position (R.A. = $14.00^\circ\pm0.05^\circ$, Decl. = $63.79^\circ\pm0.02^\circ$). The WCDA analysis yielded a larger intrinsic extension of $0.34^\circ\pm0.04^\circ$, with the best-fit position at (R.A. = $13.96^\circ\pm0.09^\circ$, Decl. = $63.92^\circ\pm0.05^\circ$).
 
For the KM2A data ($ E > 25\rm~TeV$), the TS value of the power-law model is 770.3, while it increases to 788.9 for the ECPL model. According to Wilks' theorem, the ECPL hypothesis is preferred over the power-law hypothesis with a significance of \signi{4.3}. Adding an extended source with ECPL spectrum to the GDE and CR background increases the TS value by 667.1. With 6 degrees of freedom, this corresponds to a significance of $25.37\ \sigma$. For WCDA data ($1 - 25\rm~ TeV$), the TS value of this model is 312.9. Adding an extended source to the GDE and background increases the TS value by 207.5. With 5 degrees of freedom, this corresponds to a significance of $13.72\ \sigma$. The diffuse fluxes from WCDA and KM2A are consistent with the flux of the outer region reported in the previous LHAASO paper on diffuse emission  \citep{lhaasocollaborationMeasurementUltraHighEnergyDiffuse2023}. In the following discussion, the morphology and spectrum are derived from the power-law model for WCDA and the ECPL model for KM2A. 

To study the significance of the extension of the source, we compare the $\Delta {\rm TS}$ between the extended source assumption and the point source assumption. For energies above 25 TeV, the TS value increases significantly by  {58.3} ($7.6\ \sigma$) under the extended source assumption. For energies below 25 TeV, the TS value shows an increment of {42.9} ($6.5\ \sigma$) compared to the point source assumption. The intrinsic extension of 1LHAASO J0056+6346u is further validated by comparing the distribution of $\theta^2$ for signals between LHAASO data and the models, as shown in Fig.~\ref{fig:km2a angular distribution}, where $\theta$ is the angular offset of each event to the central position of 1LHAASO J0056+6346u. A set of \grays is generated, taking into account the Spectral Energy Distribution (SED), the intrinsic source extension, and the detector Point Spread Function (PSF). The good agreement between simulation and experimental results, with {$\chi^2/ndf =8.92/10$} for WCDA and {$\chi^2/ndf =13.85/15$} for KM2A, demonstrates the correct understanding of the extension of the source.

To investigate the energy dependence of the extension of 1LHAASO J0056+6346u, we divided the KM2A data into two energy bins: 25–100 TeV and 100–400 TeV. Each bin was fitted independently to determine the source extension, with the spectral index of the source and the spectrum for the GDE fixed during the fitting. With the source extensions determined for each energy bin, we compared two models: an energy-independent extension and an energy-dependent power-law extension. The fitting results are presented in Fig.~\ref{fig:km2a angular distribution} (c). Under the first assumption, the extension was found to be \(A = 0.26^\circ \pm 0.02^\circ\). Under the second assumption, the extension parameters were obtained as \(A = 0.43^\circ \pm 0.09^\circ\) and \(B = 0.14 \pm 0.05\). The power-law model yielded a TS improvement of 6.03 over the energy-independent model, corresponding to a significance of 2.46 \(\sigma\). These results indicate a potential energy dependence of the extension, though additional data are needed to confirm this behavior.

The SED ($E^2dN\!/\!dE$) is shown in Fig.~\ref{fig:km2a angular distribution} (d). 
The SED is well described by an ECPL function, with $\chi^2/ndf$ = 12.87/9. Compared to a single power-law, the ECPL model provides a significant improvement of \signi{14.7}.
The differential flux (${\rm TeV}^{-1}\,{\rm cm}^{-2}\,{\rm s}^{-1}$) in the energy range from 1 to 400 TeV is
\begin{equation}
    \frac{dN}{dE\,dA\,dt} = (2.67\pm0.25) 
                \times 10^{-15} \left(\frac{E}{20\,{\rm TeV}}\right)^{-\alpha} 
                e^{-\frac{E}{E_{\rm cut}}}
\tag{2},
\end{equation}
where $\alpha = 1.97\pm0.10$ and $E_{\rm cut} = (55.1\pm7.2) \rm\ TeV$. The spectrum implies a cutoff around 50 TeV.  Integral flux of the ECPL spectrum above 1 TeV is $5.75\times10^{-12} \ \mathrm{erg}\,\mathrm{cm}^{-2}\,\mathrm{s}^{-1}$, corresponding to approximately 6.1\% of the flux of the Crab Nebula above 1 TeV  \citep{crab_science}.

\section{Multiwavelength observations} 

\subsection{\textit{Fermi}-LAT observation}
\label{subsec:fermi}

To explore the origin of  1LHAASO J0056+6346u, we examined the most recent \textit{Fermi}-LAT source catalog \citep[4FGL-DR4 catalog][]{4fgl-dr4}, and found that the point source 4FGL J0057.9+6326 is the closest source.
Since 4FGL J0057.9+6326 has no identified association with any known objects, we used over 15 years of \textit{Fermi}-LAT observation data (from August 4, 2008, to April 8, 2024) to re-investigate its properties in the 1--{1000} GeV energy range and to explore its potential connection to 1LHAASO J0056+6346u.

\begin{figure}[H]
    \centering
    \includegraphics[width=0.5\textwidth]{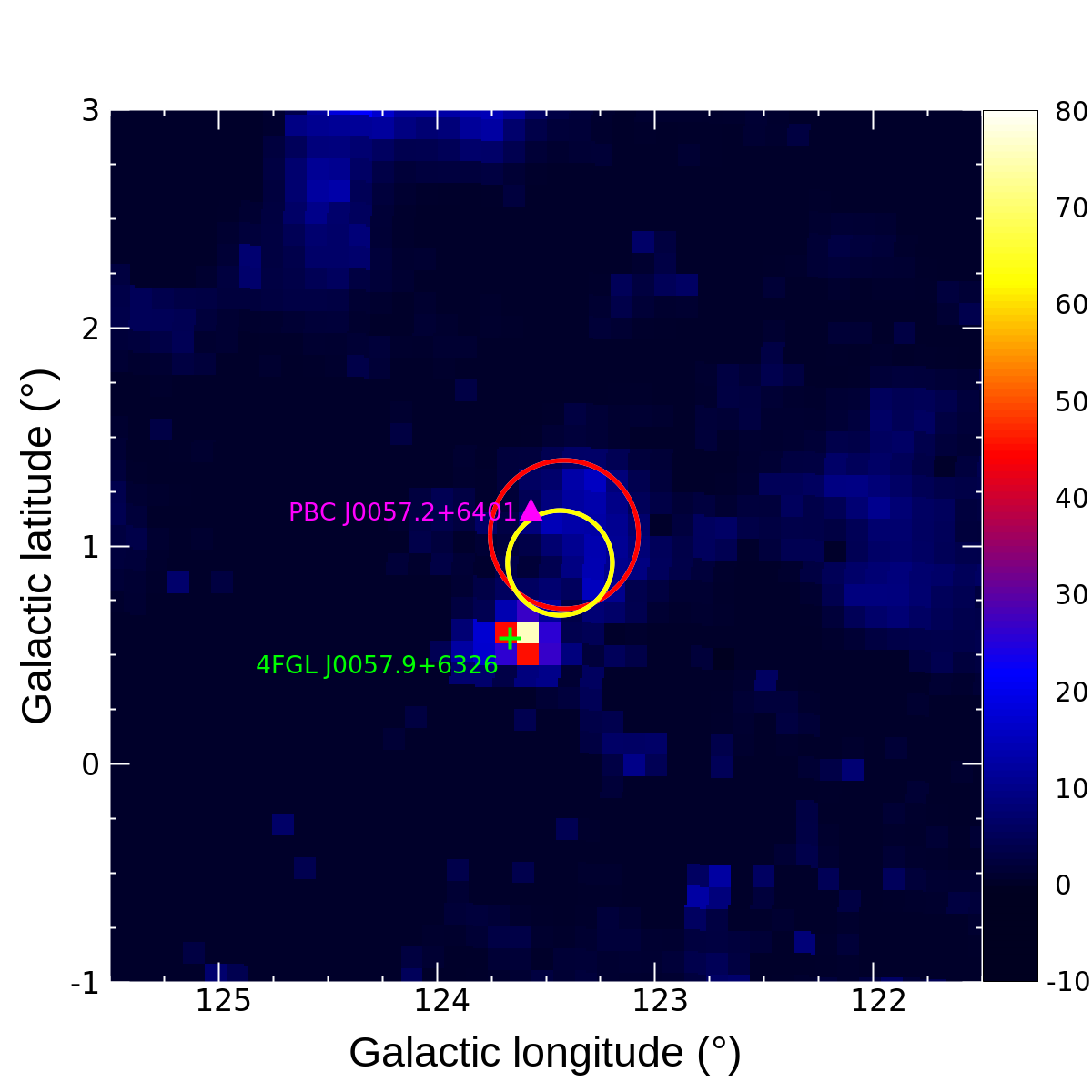}
    \caption{\textit{Fermi}-LAT residual TS map around 1LHAASO J0056+6346u. The yellow circle and red circle indicate the intrinsic extensions of 1LHAASO J0056+6346u, derived from KM2A and WCDA data, respectively. The green cross marks the position of 4FGL J0057.9+6326. The magenta triangle indicates the position of the X-ray source PBC J0057.2+6401.  
 }
    \label{fig:fermi}
\end{figure}

First, we performed a standard binned likelihood analysis using the source model generated by \texttt{ make4FGLxml.py}\footnote{\url{https://fermi.gsfc.nasa.gov/ssc/data/analysis/user/make4FGLxml.py}}, which includes point sources and diffuse emission (\texttt{gll\_iem\_v07.fits}, \texttt{iso\_P8R3\_SOURCE\_V3\_v1.txt})  based on the 4FGL-DR4 catalog  \citep{4fgl-dr4,Abdollahi_2022}. 
After subtracting all other sources in the model as background, we created a TS map (see Fig.~\ref{fig:fermi}) and found no significant deviation of the position of 4FGL J0057.9+6326 compared to the catalog position.
To test the extension of 4FGL J0057.9+6326, we replaced the point source model with a uniform disk model. The radius of the disk was varied from $0.1^{\circ}$ to $1.0^{\circ}$, with an interval of $0.1^{\circ}$.  We found no increase in the overall log-likelihood value. Based on these results, 4FGL J0057.9+6326 is more likely to be a point source.
In short, our analysis found that in the low energy range (1--1000 GeV), the position and extension of 4FGL J0057.9+6326 are consistent with the 4FGL-DR4 catalog results.
Thus far, the relation between 4FGL J0057.9+6326 and 1LHAASO J0056+6346u remains unclear.

To estimate the flux of 1LHAASO J0056+6346u in the GeV energy range, we deleted 4FGL J0057.9+6326 from the source model and added an extended source that shares the same spatial distribution derived from WCDA data, which also encompasses the region of KM2A-detected emission.
The spectral shape of this GeV source is assumed to follow a power law. For each bin, we calculated the  95$\%$ upper limit if the TS value of the source was less than 9 (corresponding to a significance of $<3\,\sigma$). The derived SED in the energy range of 1--1000\,GeV, based on the WCDA morphology, is shown as magenta data points in Fig.~\ref{fig:sedfit}. The corresponding total flux is   $(6.51\pm1.86)\times10^{-12} {\rm\  erg\,cm^{-2}\,s^{-1}}$, which can be considered an upper limit.  For comparison, the SED of 4FGL J0057.9+6326 is shown as green data points in the same figure.

Additionally, we searched for potential X-ray counterparts of 1LHAASO J0056+6346u. From the \texttt{SIMBAD} database, we found an X-ray source PBC J0057.2+6401 \citep{PBC} within the radiation region. However, PBC J0057.2+6401 is also possibly associated with an active galactic nuclei based on the positional uncertainty \citep{2017A&A...602A.124R}. The position of PBC J0057.2+6401 is indicated in Fig.~\ref{fig:fermi}.

\subsection{\gray spectral fitting}
\label{subsec:sed}
To investigate the possible radiation mechanisms of the \grays in this region, we fit the SEDs acquired from  \textit{Fermi}-LAT, WCDA, and KM2A observations with both a leptonic scenario (i.e., the Inverse Compton scattering, hereafter referred to as IC model), and a hadronic scenario (i.e., p-p inelastic collision, hereafter referred to as PP model). The fitting was performed using the \texttt{NAIMA} package\footnote{\url{http://naima.readthedocs.org/en/latest/index.html}}  \citep{naima_0}, which includes tools for performing Markov Chain Monte Carlo fitting of nonthermal radiative processes for a range of models describing the distribution of non-thermal particles.
Here, the distribution of the parent particles was assumed to be \texttt{ExponentialCutoffPowerLaw} (ECPL) or \texttt{BrokenPowerLaw} (BPL), and their formulae are listed in Table~\ref{tab:form}. Analytic parametrizations of \gray emission from p-p interactions developed by \citet{Kafexhiu2014} are implemented for the PP models, and analytic approximations for IC scattering developed by \citet{Khangulyan2014} are applied for the IC models.  In the IC models, we first consider only the Cosmic Microwave Background (CMB) for a general test, and we designate this fitting model as IC (CMB).  

\begin{table*}[]
    \centering
    \footnotesize
    \begin{threeparttable}
        \caption{Formulae for parent particle distributions used in the SED fitting process}\label{tab:sedfitfun}
        \doublerulesep 0.1pt \tabcolsep 13pt 
        \begin{tabular}{ccc}
                \toprule
                \hline
                    Name & Formula   &Free parameters   \\
                \hline  
                    ECPL & $N(E) = A (E/10\,{\rm TeV})^{-\alpha} {\exp(-(E/E_\mathrm{cut})}^{\beta}) $& A, $\alpha$, $\beta$, $ E_\mathrm{cut}$ \\
                    BPL &  $N(E) =\begin{cases} A(E/10\,{\rm TeV})^{-\alpha_1} & \mbox{: }E<E_\mathrm{b} \\ A(E_\mathrm{b}/10\,{\rm TeV})^{(\alpha_2-\alpha_1)}(E/10\,{\rm TeV})^{-\alpha_2} & \mbox{: }E>E_\mathrm{b} \end{cases}$&  A, $\alpha_1$, $\alpha_2$, $E_\mathrm{b}$\\
                \hline
                \bottomrule
        \end{tabular}
        \label{tab:form}
    \end{threeparttable}

    \hskip 10pt

    \footnotesize
  
    \begin{threeparttable}
        \caption{The fitting results of different radiation models \tnote{1)}}
        \doublerulesep 0.1pt \tabcolsep 13pt 
            \begin{tabular}{cccccccc}   
                \toprule
                \hline
                    Model & Formula &$\alpha$/$\alpha_1$ &$\beta$ / $\alpha_2$  & $E_{\rm cut}$/$E_{\rm b}$ (TeV)   &$W_{\rm p}$/$W_{\rm e}$ (erg)\tnote{2)} & MLL \tnote{3)}\\         
                \hline             
               {PP} &ECPL& $1.92_{-0.08}^{+0.08}$ & $3.03_{-0.96}^{+1.20}$ & $442_{-53}^{+48}$& $(1.44_{-0.27}^{+0.38})\times10^{49} $  &$-6.39$ \\
              ( $n_{\rm H}=1\,{\rm cm}^{-3}$ )  & BPL &  $1.83_{-0.08}^{+0.32}$ & $4.55_{-0.56}^{+0.28}$ & $253_{-55}^{+33}$ &$(1.30_{-1.2}^{+0.35})\times10^{49}$ &$-6.65$  \\
                \hline
        {IC }&ECPL & $2.04_{-0.44}^{+0.32}$ & $1.08_{-0.23}^{+0.34}$ & $90_{-43}^{+52}$&$(0.47_{-0.35}^{+3.33})\times10^{47}$  &$-6.43$ \\
          (CMB)       & BPL & $2.60_{-0.21}^{+0.10}$ & $4.80_{-0.22}^{+0.13}$ &$119_{-16}^{+12}$ &$(1.79_{-1.50}^{+2.79})\times10^{48}$  & $-8.23$ \\
                       \hline
    	    {Syn+IC} &ECPL & $2.34_{-0.37}^{+0.22}$ & $1.41_{-0.36}^{+0.52}$ &$144_{-60}^{+51}$ &$(2.17_{-1.95}^{+9.88})\times10^{46}$  & $-6.43$\\
	    	  (CMB, $B=10\,{\rm \mu G}$)   & BPL  & $2.55_{-0.25}^{+0.12}$ & $4.72_{-0.26}^{+0.20}$ & $113_{-17}^{+15}$&$(9.07_{-8.18}^{+1.94})\times10^{46}$  & $-8.28$\\           \hline
      	    {Syn+IC }  &ECPL & $2.51_{-0.01}^{+0.01}$ & $2.63_{-0.65}^{+1.23}$ &$202_{-13}^{+17}$ &$(2.40_{-0.12}^{+0.11})\times10^{46}$  & $-14.73$\\
	    	(G124, $B=3\,{\rm \mu G}$)     & BPL  & $2.51_{-0.01}^{+0.01}$ & $4.70_{-0.25}^{+0.20}$ & $111_{-11}^{+12}$&$(2.41_{-0.12}^{+0.12})\times10^{46}$  & $-14.15$\\ 
                  \hline
      {Syn+IC } &ECPL & $2.28_{-0.02}^{+0.02}$ & $1.82_{-0.35}^{+0.61}$ &$163_{-19}^{+18}$ &$(4.09_{-0.18}^{+0.19})\times10^{45}$  &$-16.07$ \\
	    	(G124, $B=10\,{\rm \mu G}$)             & BPL  & $2.29_{-0.01}^{+0.01}$ & $4.65_{-0.25}^{+0.22}$ &$99_{-9}^{+9}$ &$(4.01_{-0.18}^{+0.17})\times10^{45}$  &$-16.60$ \\      
                 \hline                        
                 \hline
                \bottomrule
            \end{tabular}
            \label{tab:fiterr}              
            \begin{tablenotes}
                \item[1)] The source distance $d$ is assumed to be 2 kpc for PP and IC models, and 553\,pc for all Syn+IC models. 
                \item[2)] The required energy budget above 1  GeV.
                \item[3)] Maximum of log-likelihood.
            \end{tablenotes}
    \end{threeparttable}
\end{table*}

Next, to test whether the X-ray source PBC J0057.2+6401 is related to the LHAASO source, we used synchrotron emission (assuming $B = 10\,\mu$G) and inverse Compton (IC) emission (considering only the CMB) from a single electron distribution to jointly fit the flux of PBC J0057.2+6401 and the $\gamma$-ray data. We refer to this test model as Syn+IC (CMB). 
To further investigate the relationship among LHAASO J0056+6346u, PBC J0057.2+6401, and the radio source G124.0+1.4, we applied Syn+IC models to fit the radio emission of G124.0+1.4 (using data from Table 1 of \citet{Chen2023AJ}), along with the X-ray and $\gamma$-ray data. In addition to the CMB, we included the far-infrared emission and starlight at the positions of the candidate accelerators, G124.0+1.4. These local far-infrared and optical emission is derived from the local interstellar radiation field calculated by \citet{popescu17}. We refer to this test model as Syn+IC (G124).

\begin{figure*}[]
   \centering
 \includegraphics[width=0.95\textwidth]{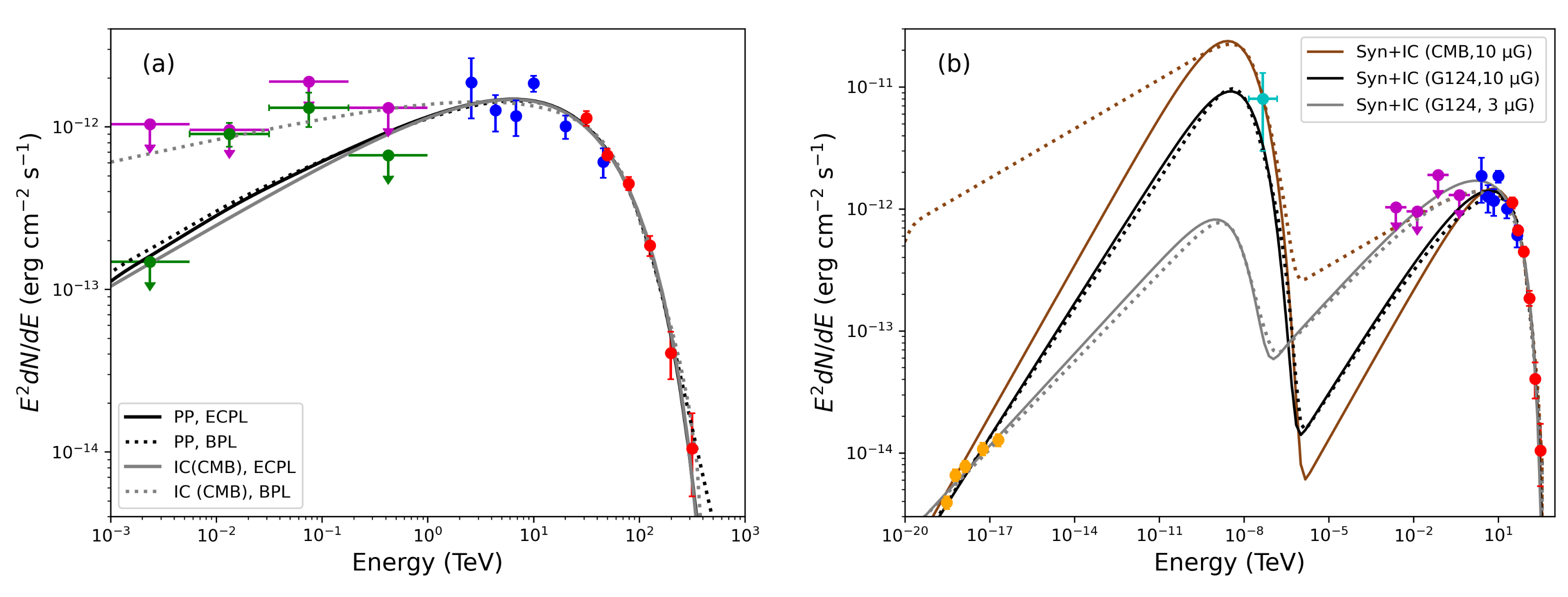}
    \caption{(a) SED fitting results for the PP and IC models based on spectral data from WCDA (blue), KM2A (red), and \textit{Fermi}-LAT (magenta) observations, with 4FGL J0057.9+6326 fluxes shown as green points for reference. 
    (b) SED fitting results of Syn+IC models.  
    The orange data points represent the radio flux of G124.0+1.4 and the cyan data points indicate the flux of the X-ray source PBC J0057.2+6401. 
   For lines of the same color, solid lines represent results using the ECPL distribution, whereas dotted lines correspond to results using the BPL distribution. 
 The gas density, background photons, or the magnetic field applied for different models are indicated in the legend.   See Sec.~\ref{subsec:sed} for further details. }
    \label{fig:sedfit}
\end{figure*}

The spectral parameters of the parent particles and the required energy budget i.e., $W_{\rm p}\,(>1 \rm{GeV})$ and $W_{\rm e}\,(>1 \rm{GeV})$ assuming different source distances $d$ are listed in Table~\ref{tab:fiterr}. The \gray spectra generated by the best-fit results of different radiation models are shown in Fig.~\ref{fig:sedfit}. 

Moreover, given the same $\gamma$-ray spectra and background photons, the required energy budget for CR protons in PP models ($W_{\rm p} \propto d^2/n_{\rm H}$) and for electrons in IC models ($W_{\rm e} \propto d^2$) depends on the source distance $d$ and gas density $n_{\rm H}$. Therefore, with different assumptions of source distance and gas density, the required $W_{\rm p}$ or $W_{\rm e}$ within the emission region will vary accordingly. Further discussions are presented in Sec.~\ref{sec:dis}.

As shown in Table~\ref{tab:fiterr} and Fig.~\ref{fig:sedfit}, both the hadronic and leptonic scenarios can adequately fit the spectrum, as judged by the maximum log-likelihood (MLL). However, the current \gray data alone cannot distinguish between these two radiation mechanisms. If the \grays are produced via the IC process, the emission is likely associated with a pulsar, suggesting it might originate from a PWN or a pulsar halo. On the other hand, in the hadronic scenario, the presence of dense gas is required to interact with CR protons accelerated by nearby sources. To further investigate the origin of 1LHAASO J0056+6346u, we examined the gas content in the vicinity of the Very High Energy (VHE) \gray emission region.

\subsection{Study of the gas content} 
\label{subsec:gas}

Previous research has found two gas bubble/shell structures around this region.
\citet{Chen2023AJ} discovered a molecular gas bubble within the $V_{\rm lsr}$ range of $-1\,\rm{km\, s^{-1}}$ to $7\, \rm{km\, s^{-1}}$ that is possibly blown by the progenitor of G124.0+1.4,  and  \citet{Cazzolato2003} also reported a partial shell of the molecular cloud at the velocity around $-48\, \rm{km\, s^{-1}}$ that is likely associated with  OB association Cas OB7. 
However, the angular sizes of these gas bubbles are much larger than that of 1LHAASO J0056+6346u, which obscures the relation between the UHE \grays and these bubbles.

\begin{figure*}[ht!]
    \centering
    \includegraphics[width=1\textwidth]{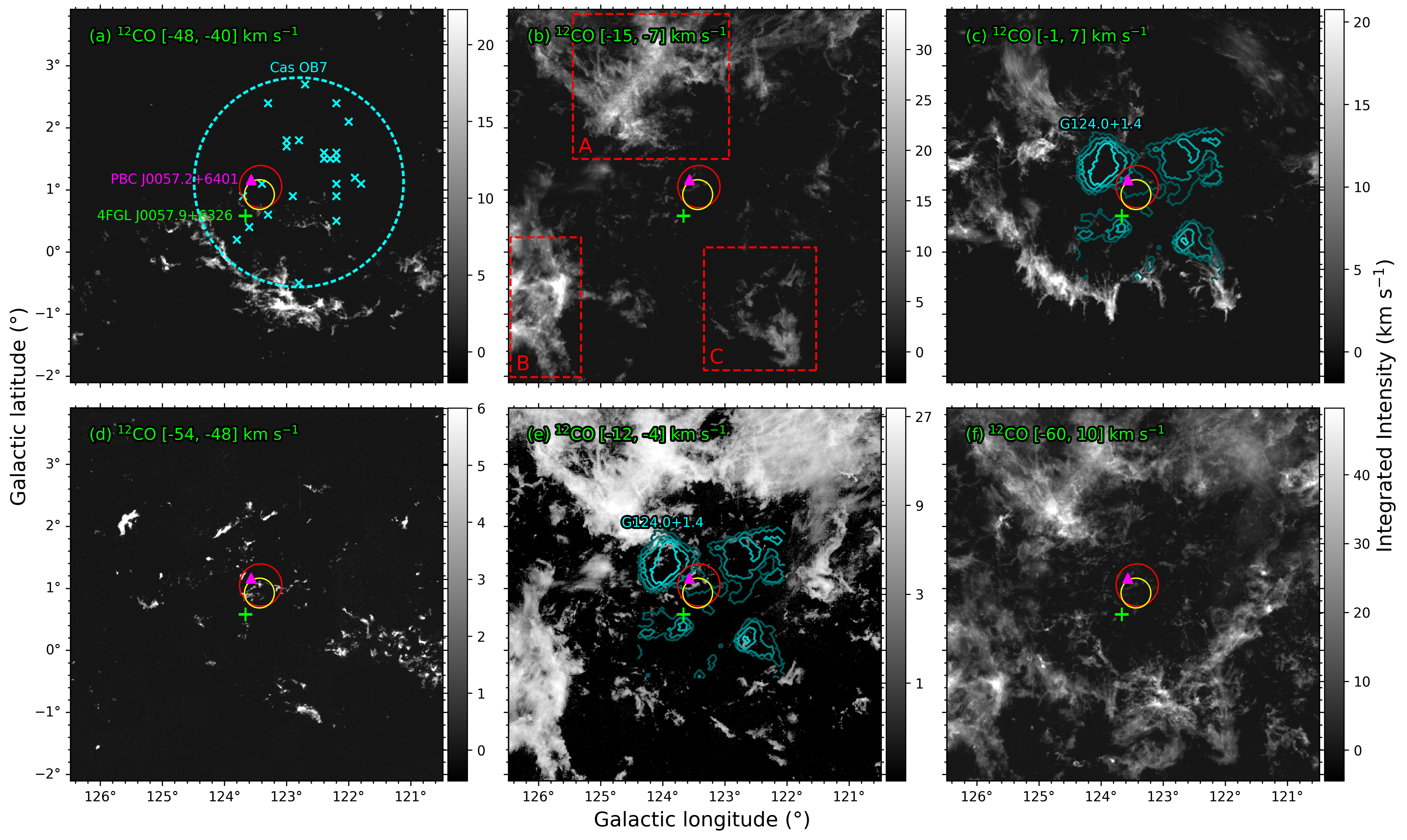}
    \caption{
    Integrated intensity maps of $^{12}$CO ($J = 1 \to 0$) line. The  yellow and red circles represent the intrinsic extensions of 1LHAASO J0056+6346u, derived from KM2A and WCDA data, respectively. The green cross marks the position of 4FGL J0057.9+6326.  The magenta triangle marks the X-ray source PBC J0057.2+6401. (a)  $^{12}$CO integrated intensity map over the velocity range $[-48, -40] \rm\ km\ s^{-1}$. The cyan dashed circle shows the Cas OB7 association. The cyan diagonal cross symbols show the positions of stars in Cas OB7. (b) $[-15, -7] \rm\ km\ s^{-1}$. The red region represents the area used for distance measurement. See Supplementary Materials.  (c) $[-1, 7] \rm\ km\ s^{-1}$. The contour shows the radio emission in the Urumqi 6 cm survey  \citep{Urumqi_image}. The bright region in the upper left shows the radio emission from the SNR candidate G124.0+1.4. (d) $[-54, -48] \rm\ km\ s^{-1}$, which corresponds to one of the velocity ranges shown in Fig.~\ref{fig:12CO_line}. (e) $[-12, -4] \rm\ km\ s^{-1}$,  which corresponds to the other velocity range shown in Fig.~\ref{fig:12CO_line}. The color axis uses a logarithmic scale for clarity.  (f) $[-60, 10] \rm\ km\ s^{-1}$.  The integrated intensity map covers the entire velocity range we investigated.
    }
    \label{fig:multiwavelength}
\end{figure*}

In this work, we conducted a detailed investigation of the gas content in the region surrounding 1LHAASO J0056+6346u using  observation data of $\rm ^{12}CO$ and $\rm ^{13}CO$ lines from the Milky Way Imaging Scroll Painting (MWISP)\footnote{\url{http://www.radioast.nsdc.cn/mwisp.php}} survey. The MWISP survey is an unbiased and sensitive CO(1–0) multiline survey toward the northern Galactic plane, conducted with the 13.7 m millimeter telescope of the Purple Mountain Observatory (PMO) \citep{MWISP_su, MWISP_sun}. 
Complementary HI line data were obtained from the Canadian Galactic Plane Survey (CGPS)\footnote{\url{https://www.cadc-ccda.hia-iha.nrc-cnrc.gc.ca/en/cgps/}} \citep{CGPS1, CGPS2}, which carried out a high-resolution survey of the atomic hydrogen and radio continuum emission from our Milky Way galaxy. The integrated intensity maps of $\rm ^{12}CO$ are presented in Fig.~\ref{fig:multiwavelength}, while the corresponding maps of $\rm ^{13}CO$ and HI spectral lines are shown in Fig.~S1 in Supplementary Materials.

As shown in Fig.~\ref{fig:multiwavelength} and Fig.~S1, in addition to the two bubble/shell structures mentioned earlier, we identified a potential bubble structure in the velocity range of $-15\,\rm{km\, s^{-1}}$ to $-7\, \rm{km\, s^{-1}}$ in the $\rm ^{12}CO$ and $\rm ^{13}CO$ data.  Enhanced HI emission is also observed in this region, and it appears to be enclosed by the bubble-like structure visible in the CO data (Fig.~S1; see also Fig.~5  in \citet{Cazzolato2003} for larger and clearer view). A possible explanation for this phenomenon is that concentrated stellar winds or SNRs have dissociated molecular hydrogen, leading to the observed enhancement in HI emission. Similar structures were also identified in the IRAS (HIRES) images at 12, 25, 60, and 100 $\mu$m \citep{IRAS} (see Fig.~1 in \citet{Cazzolato2003}) and in the PLANCK frequency maps from the High Frequency Instrument \citep{PLANCK_HFI}. The integrated intensity map in the entire velocity range from $-60$ to $10\, \mathrm{km\,s^{-1}}$ which we investigated, is also plotted in Fig.~\ref{fig:multiwavelength}.

Using the MWISP CO data and Gaia DR3 \citep{gaia_mission, gaia_dr3_summary, gaia_edr3_am, gaia_ap}, we estimated the distance to the potential bubble structure at the velocity range of $-15$ to $-7\, \mathrm{km \, s^{-1}}$ following the method described in \citet{distance_fit_method}. The derived distance is approximately 800–900 pc, with details provided in the Supplementary Materials.

\begin{figure}[H]
    \centering
    \includegraphics[width=0.5\textwidth]{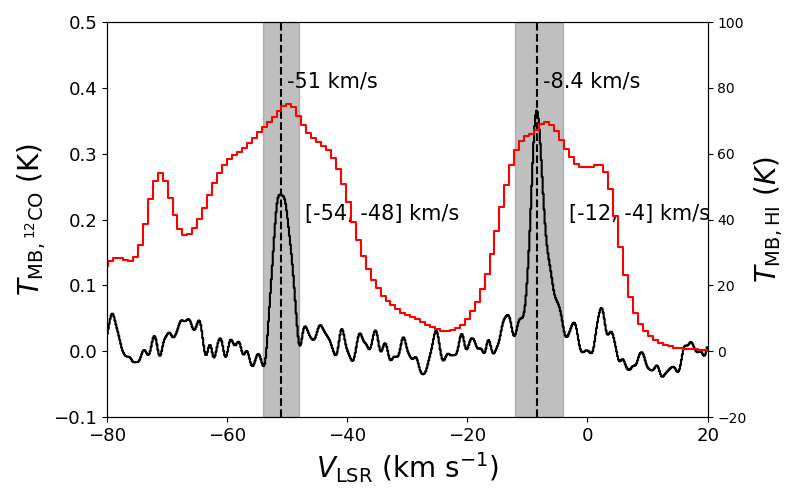}
    \caption{ Mean intensities of the $^{12}$CO ($J = 1 \to 0$) (black) and HI (red) spectral lines within the WCDA emission region.}
    \label{fig:12CO_line}
\end{figure}

Furthermore, we investigated the gas content within the radiation region, defined as a circular area with an angular radius of $0.34^{\circ}$, based on the intrinsic extension determined by WCDA data. Significant gas components were detected in the $\rm ^{12}CO$ data (Fig.~\ref{fig:12CO_line}), whereas no significant velocity components were detected in the $\rm ^{13}CO$ data. Two distinct velocity components were detected: one at $-54$ to $-48 \, \mathrm{km \, s^{-1}}$, and the other at $-12$ to $-4 \, \mathrm{km \, s^{-1}}$. The integrated intensity maps corresponding to these ranges are also shown in Fig.~\ref{fig:multiwavelength}. 

Using the $\rm ^{12}CO$ and HI data for these two velocity ranges, we calculated the corresponding gas content. The $\rm H_2$ column density was estimated using:
\begin{equation}
\centering
    N_{\rm H_2} = X\,W_{^{12}\rm CO}=X \int{T_{mb, ^{12}\rm CO}}\ d\,V
\tag{3}, 
\end{equation} where $T_{mb, ^{12}\rm CO}$ is the main-beam temperature of $\rm ^{12}CO$ as a function of velocity. Here, we used a mean CO-to-$\rm H_2$ conversion factor  $ X= 1.8 \times 10^{20} \rm\ cm^{-2}\ K^{-1}\ km^{-1} s $ \citep{Dame_12COtoH2}. The mean $\rm H_2$ column density within the WCDA emission region is $1.46 \times 10^{20} \, \mathrm{cm}^{-2}$ for $-54$ to $-48 \, \mathrm{km \, s^{-1}}$ and $2.22 \times 10^{20} \, \mathrm{cm}^{-2}$ for $-12$ to $-4 \, \mathrm{km \, s^{-1}}$.

The HI column density was estimated using \citep{wilson2013}:
\begin{equation}
    N_{\rm HI} = 1.8 \times 10^{18} \int{T_{mb, \rm HI}}\ d\,V
\tag{4}.
\end{equation}
As shown in Fig.~\ref{fig:12CO_line}, the range of the HI spectral line is significantly broader than that of 
 $\rm ^{12}CO$, complicating the definition of a common integration interval. If we conservatively select the same velocity range as for the $\rm ^{12}CO$ emission, for $-54$ to $-48 \, \mathrm{km \, s^{-1}}$, the mean {\rm HI} column density is $9.58 \times 10^{20} \, \mathrm{cm}^{-2}$, and for $-12$ to $-4 \, \mathrm{km \, s^{-1}}$, the mean {\rm HI} column density is $1.07 \times 10^{21} \, \mathrm{cm}^{-2}$.

Therefore, the mean hydrogen column density $N_{\rm H} = 2N_{\rm H_2} + N_{\rm HI}$ within the WCDA emission region in the velocity ranges of $-54$ to $-48 \, \mathrm{km \, s^{-1}}$ and $-12$ to $-4 \, \mathrm{km \, s^{-1}}$ is estimated to be $1.25 \times 10^{21} \, \mathrm{cm}^{-2}$ and $1.51 \times 10^{21} \, \mathrm{cm}^{-2}$, respectively.

Then we estimated the kinematic distance using the \texttt{kd} package\footnote{\url{https://zenodo.org/records/1166001}} \citep{Wenger2018}, which applies a Monte Carlo method to compute kinematic distances and their associated uncertainties. This method incorporates the rotation curve and updated solar motion parameters from \citet{reid2014}. The estimated distances are $4.05^{+0.40}_{-0.32}\ \rm kpc$ ($V_{\rm LSR} = -51 \rm\ km\ s^{-1}$) and $623^{+58}_{-51}\rm\ pc$ ($V_{\rm LSR} = -8.4 \rm\ km\ s^{-1}$), respectively. However, according to the ``two-armed spiral shock'' (TASS) model \citep{Roberts1972}, as discussed by \cite{Cazzolato2003}, the gas content in the velocity range of $-54$ to $-48 \, \mathrm{km\,s^{-1}}$ is also possibly located near Cas OB7, which is approximately $2\,$kpc away.  

Finally, based on the parameters derived above, we calculated the total gas masses within the radiation region, $M_{\rm H}=N_{\rm H}\times A$,  in which $A=\pi [d\tan(r_{39})]^{2}$, $d$ is the gas distance and $r_{39}=0.34^{\circ}$ is the \gray source extension derived from WCDA data. 
For simplicity,  the average number density $n_{\rm H}$ is estimated assuming that the gas {in certain velocity range} is within a sphere with a radius of  $\sim d\tan(r_{39})$. For the gas in velocity range $-54$ to $-48 \, \mathrm{km\,s^{-1}}$  , the mass is $\sim1.7\times10^4\,{\rm M_{\odot}}$ and  the density is $\sim 13\,{\rm cm}^{-3}$  assuming a distance of $\sim 4$\,kpc or $\sim4.3\times10^3\,{\rm M_{\odot}}$ and  $\sim 26\,{\rm cm}^{-3}$ assuming a distance of $\sim 2$\,kpc. For the gas in velocity range $-12$ to $-4 \, \mathrm{km\,s^{-1}}$, the mass is $\sim490\,{\rm M_{\odot}}$ and  the density is $\sim 102\,{\rm cm}^{-3}$.

\section{Discussion on the origin of 1LHAASO J0056+6346u}
\label{sec:dis}

To find the origin of 1LHAASO J0056+6346u, we searched for possible CR accelerators in its vicinity. We found neither known pulsar in the ATNF pulsar catalog \citep{atnf} \footnote{ \url{http://www.atnf.csiro.au/research/pulsar/psrcat}}, nor SNR in Green's catalog \citep{green_catalog}.  However, in the line of sight, there is a recently discovered SNR candidate, G124.0+1.4 \citep{Chen2023AJ}, which partially overlaps with the gamma-ray emission region, and also an OB association, Cas OB7 \citep{Humphreys1976ApJ}, which covers the whole \gray emission region. Both could serve as potential CR accelerators that might contribute to the observed UHE $\gamma$-ray emission.
Additionally, a PWN or pulsar halo associated with an unknown pulsar could also explain the $\gamma$-ray emission. 
We discuss these three potential accelerator scenarios in the following sections.

\subsection{A PWN or pulsar halo from an unknown pulsar}

Although 1LHAASO J0056+6346u does not overlap with any known pulsar in the line of sight, we cannot formally rule out the possibility that the detected UHE \gray emission is from the PWN of an off-beamed pulsar.  Considering the effects of rapid cooling of electrons above $\sim$100 TeV, the energy-dependent morphology of 1LHAASO J0056+6346u also favors a leptonic origin. Moreover, the energy budget of electrons ($W_e$) can be further reduced by considering the local interstellar radiation field  for IC emission.  As shown by the IC model fitting results (see Table~\ref{tab:fiterr} in Sec.~\ref{subsec:sed}),
the cutoff in the \gray spectrum is likely a result of the cutoff at $\sim 100$\,TeV in the parent electron population. 
The very large index break from the BPL distributions, i.e., $(\alpha_2-\alpha_1)>2$, is inconsistent with a cooling break in the electron energy distribution. Thus, the ECPL models, in which $\alpha\sim2$, $\beta\sim1$ and $E_{\rm cut}\sim 100$\,TeV, are favored for describing the parent electron distributions.

Moreover, along the line of sight, PBC J0057.2+6401 lies at the edge of G124.0+1.4 and within the emission region of LHAASO J0056+6346u. The X-ray emission from PBC J0057.2+6401 may be extended but is classified as a point source possibly because of the  the influence of thermal X-rays and the limited sensitivity of the telescope. Therefore, it is possible that this hard X-ray source is associated with the UHE $\gamma$-ray emission. Assuming a magnetic field strength of $B = 10\,\mu$G, the synchrotron and IC emissions from a single population of electrons (Syn+IC model) with an ECPL distribution can also fit the X-ray and $\gamma$-ray data very well, as shown by the brown lines in Fig.~\ref{fig:sedfit}~(b). This suggests that the $\gamma$-ray and X-ray emissions may share the same origin, which is likely to be a PWN. However, as mentioned earlier, PBC J0057.2+6401 is possibly associated with an AGN \citep{2017A&A...602A.124R}. Thus, to confirm the connection between the TeV radiation and PBC J0057.2+6401, more evidence is needed.

\subsection{SNR candidate G124.0+1.4}

SNRs are considered primary CR accelerators in the Milky Way \citep{Blasi13}. Despite ongoing debates on whether SNRs are PeVatrons \citep[e.g.,][]{Cristofari2020}, evidence from \gray observations has revealed that SNRs, especially young, energetic ones, can accelerate both protons and electrons to very high energies. The hard \gray spectrum of 1LHAASO J0056+6346u resembles that of other young SNRs, such as RXJ1713-3946 \citep{Celli2019,hess_1713} and Vela Junior \citep{hess_velaj}.  Thus, we first examine the possibility that the UHE \grays originated from the particles accelerated by the SNR candidate G124.0+1.4, which partially overlaps with 1LHAASO J0056+6346u in the line of sight. 

According to \cite{Chen2023AJ}, the distance of the gas shell associated with G124.0+1.4 is $553^{+8}_{-9}\rm\ pc$ and the radius of G124.0+1.4 is approximately 5 pc. According to the evolution theories of SNRs \citep[e.g.,][]{1999ApJS..120..299T,1972ARA&A..10..129W}, G124.0+1.4 is very likely a relatively young remnant that has recently entered Sedov-Taylor phase \citep{Sedov1959}.  Thus, G124.0+1.4 could be a powerful CR accelerator supposing it is a real SNR. Because of the uncertainty of the gas distance estimation, the gas in the velocity range of $-12$ to $-4 \, \mathrm{km\,s^{-1}}$ , whose distance is estimated as $623^{+58}_{-51}\rm\ pc$  (see Sec.~\ref{subsec:gas} for details), 
could be closer to the SNR candidate (with a distance of less than 20 pc) than the gas bubble blown by its hypothetic progenitor.
Consequently, the gas could be illuminated by escaping CRs from the SNR candidate that have not reached the gas bubble. 
Based on the  density ($n_{\rm H}\sim102\,{\rm cm}^{-3}$)  and distance of the gas  (623\,pc ), the total energy ($>1$\,GeV) of the  CR protons  interacting with the gas (within a radius of $\sim3.7$\,pc) is estimated to be $\sim 1.3\times10^{46}$\,erg.  
If the SNR-accelerated CRs are uniformly distributed within a sphere of radius 20\,pc, then the total proton energy is  $\sim2\times10^{48}$\,erg, which is much less than $1\%$ of the canonical explosion energy of an ordinary SNR. 

Additionally, IC emission from SNR-accelerated electrons scattering low-energy background photons could also be the origin of these UHE $\gamma$-rays. Given the closer distance of G124.0+1.4, the required energy for the SNR-accelerated electrons $W_{\rm e}$ to produce the emission is less than  $\sim1.3\times10^{47}$\,erg,  which is derived from the acquired $W_{\rm e}$ of IC (CMB, BPL) model (see Table.~2 for details). 
Furthermore, when including radio emission from G124.0+1.4 and X-ray emission from PBC J0057.2+6401, the Syn+IC model for a single electron population (assuming $B = 10\,\mu\mathrm{G}$) provides only a marginal fit to the SED data (gray and black lines in Fig.~\ref{fig:sedfit} (b)).

\subsection{OB association Cas OB7}

Recently, the young massive star clusters have been recognized as a population of \gray emitters and possible CR accelerators \citep{aharonian20}. In this regard, Cas OB7, containing over 20 OB stars \citep{Cazzolato2003}, also has the potential to accelerate CRs.  The distance to Cas OB7 is estimated to be 1.8 kpc in \citet{casob7_distance1} and 2.3 kpc in \citet{Humphreys1976ApJ}, respectively. 

The difficulty in this scenario is that, unlike other \gray emitting stellar clusters, the size of the \gray emission here is far smaller than the OB association itself.  
If the CRs are accelerated by the collective winds inside the cluster \citep{vieu22}, the emission region can be compact.  Moreover, if the gas in the velocity range of $-54$ to $-48 \, \mathrm{km\,s^{-1}}$ is indeed closer to the acceleration site than the gas shell associated with Cas OB7, the angular size of the \gray emission could be consistent with the bombarded gas  rather than the outer shell that might not be bright enough to be detected. Given the total mass of the gas  (assuming a distance of 2\,kpc), the required $W_{\rm p}\, (>1\,\rm {GeV})$ to illuminate this  gas  is $\sim 5.5\times 10^{47}$\,erg, which will further alleviate the energy budget and efficiency problem. 

To further investigate the origin of the $\gamma$-ray emission, we searched for OB stars in the WCDA radiation region using the \texttt{Gaia gold sample oba stars} table \citep{gaia_oba} but did not find any O-type stars. Within 1.5 kpc, there are only four B-type stars, which are likely located in the Local Arm. Beyond 1.5 kpc, up to a maximum distance of 7 kpc, there are 81 B-type stars, likely located in the Perseus Arm or even farther regions. However, the distribution of these stars is diffuse. We found no clear evidence of clustering. Therefore, the diffuse distribution of these stars suggests this scenario is currently disfavored. 

Recently, LHAASO has detected several new unidentified UHE sources, such as LHAASO J0341+5258  \citep{LHAASOJ0341} and LHAASO J2108+5157 \citep{LHAASOJ2108}. All these sources reveal a similar spectral shape: an upper limit in the GeV band, a hard (power-law index $<-2$ ) spectral shape in the TeV band, and a clear cutoff at dozens of TeV. This similar spectral shape may suggest a possible similar origin. Such a hard spectral shape in the TeV band can be formed in a scenario in which dense clouds are illuminated by the energy-dependent propagation of accelerated particles \citep{mitchell24}. In this case, the \gray emission should be spatially correlate with molecular clouds.  On the other hand, although no pulsar has yet been detected in the vicinity, we cannot formally rule out the possibility that the \gray emission from 1LHAASO J0056+6346u is from PWN or TeV halo associated with some pulsar whose pulsed emission does not beam toward Earth. In both cases, multiwavelength studies, especially those focusing on the gas distribution and X-ray observations, would be crucial to understand the origin of these UHE \gray emissions. 

\section{Conclusions}

In this work, we conducted more in-depth data analysis and discussion on the UHE \gray source 1LHAASO J0056+6346u. The source reveals an energy-dependent morphology with an average extension of $\sim0.3^{\circ}$ and a significantly curved \gray spectrum {with an exponential cutoff at $\sim55$\,TeV}.  
Since both leptonic (IC scattering) and hadronic (pp inelastic collisions) scenarios can explain the SED, we then conducted a deep investigation of the gas distribution around 1LHAASO J0056+6346u to identify potential target gas. We found that the \gray emission spatially coincides with gas in two velocity ranges, which might be illuminated by CR protons from nearby sources. 
Possible CR accelerator candidates include the SNR candidate G124.0+1.4 and the OB association Cas OB7. 
We found that G124.0+1.4 has the potential to accelerate CR protons to the energies of hundreds of TeV. However, we cannot confirm this association due to large uncertainties in the estimated gas distances.  In contrast, the Cas OB7 origin scenario is disfavored due to the discrepancies in the spatial distributions of the early type stars, gas, and the detected \gray emissions.  Moreover, a PWN powered by an unknown pulsar could naturally explain both the spatial and spectral properties of 1LHAASO J0056+6346u. 
Future multiwavelength studies, especially in the hard X-ray band, such as observations from the Einstein Probe (EP) mission \citep{EP}, may be crucial to pin down the radiation mechanism. Deeper studies of pulsars and gas properties in this region are also required for a definite association of 1LHAASO J0056+6346u  with other astrophysical counterparts.

\section*{Acknowledgements}

We would like to thank all staff members working year-round at the LHAASO site above 4400 meters above sea level to maintain the detector and keep the water recycling system, electricity power supply, and other experiment components operating smoothly. We are grateful to Chengdu Management Committee of Tianfu New Area for the constant financial support for research with LHAASO data. We appreciate the computing and data service support provided by the National High Energy Physics Data Center for the data analysis in this paper. This research work is supported by the following grants: The National Natural Science Foundation of China No.12393854, No.12393851, No.12393852, No.12393853, No.12205314, No.12105301, No.12305120, No.12261160362, No.12105294, No.U1931201, No.12375107, No.12173039, the Department of Science and Technology of Sichuan Province, China No.24NSFSC2319, No.2024NSFSC0449, Project for Young Scientists in Basic Research of Chinese Academy of Sciences No.YSBR-061, and in Thailand by the National Science and Technology Development Agency (NSTDA) and the National Research Council of Thailand (NRCT) under the High-Potential Research Team Grant Program
(N42A650868).

This research also made use of the data from the Milky Way Imaging Scroll Painting (MWISP) project, which is a multi-line survey in 12CO/13CO/C18O along the northern galactic plane with PMO-13.7m telescope. We are grateful to all the members of the MWISP working group, particularly the staff members at PMO-13.7m telescope, for their long-term support. MWISP was sponsored by National Key R\&D Program of China with grants 2023YFA1608000 \& 2017YFA0402701 and by CAS Key Research Program of Frontier Sciences with grant QYZDJ-SSW-SLH047.

The research presented in this paper has used data from the Canadian Galactic Plane Survey, a Canadian project with international partners, supported by the Natural Sciences and Engineering Research Council.

This work has made use of data from the European Space Agency (ESA) mission
{\it Gaia} (\url{https://www.cosmos.esa.int/gaia}), processed by the {\it Gaia}
Data Processing and Analysis Consortium (DPAC,
\url{https://www.cosmos.esa.int/web/gaia/dpac/consortium}). Funding for the DPAC
has been provided by national institutions, in particular the institutions
participating in the {\it Gaia} Multilateral Agreement.

\InterestConflict{The authors declare that they have no conflict of interest.}




\bibliographystyle{aasjournal}
\bibliography{main}





\renewcommand{\thefigure}{S\arabic{figure}}
\setcounter{figure}{0}
\section*{Supplementary Materials for CO and HI Data Analysis}

\begin{figure*}[p]
    \centering
    \includegraphics[width=1\textwidth]{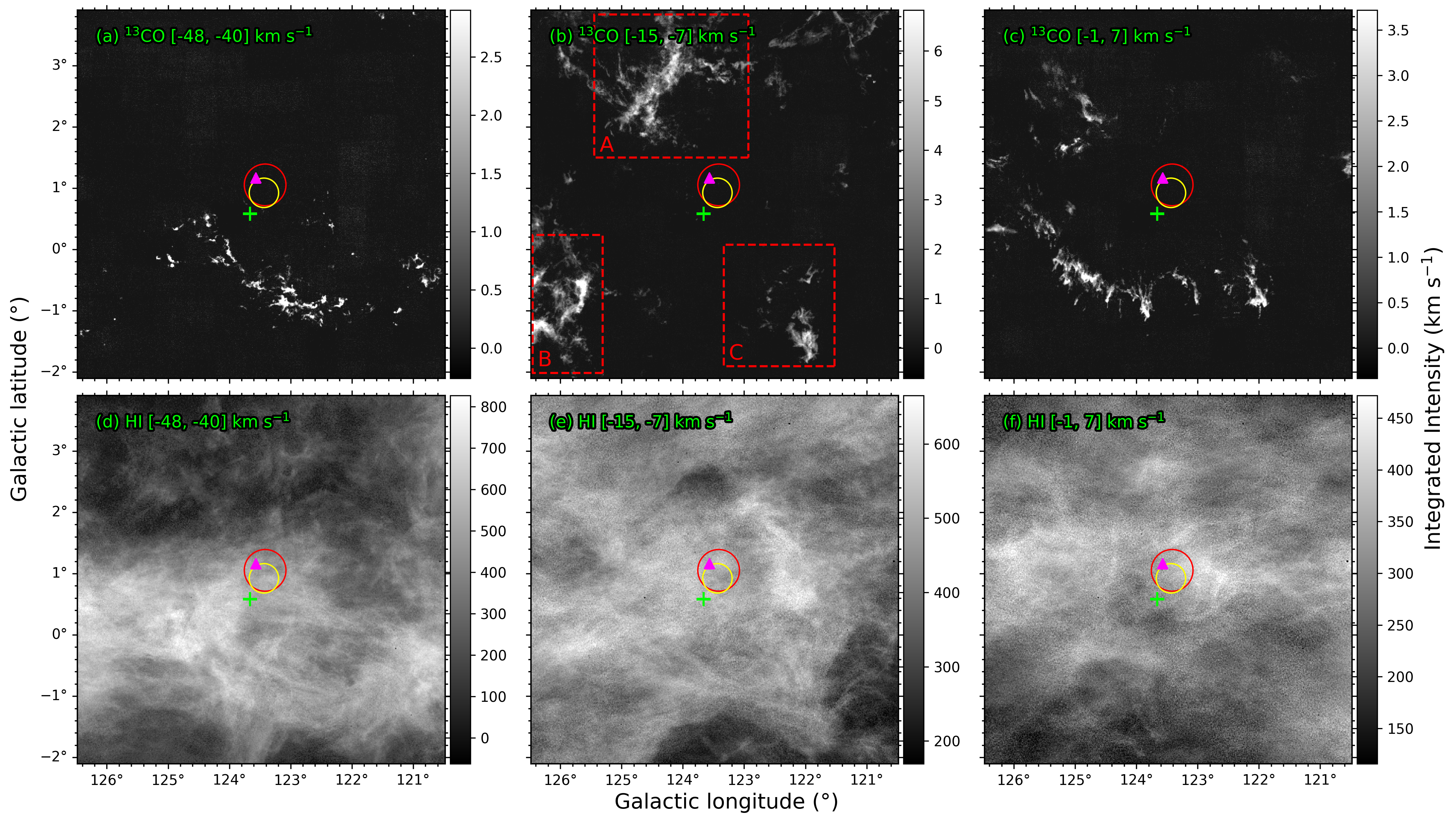}
    \caption{Integrated Intensity maps of $^{13}$CO ($J = 1 \to 0$) and HI $21\,$cm spectral lines. The markers and regions denote the same features as in Fig.~ \ref{fig:multiwavelength}. (a--c)  $^{13}$CO integrated intensity map in the velocity range $[-48, -40] \rm\ km\ s^{-1}$, $[-15, -7] \rm\ km\ s^{-1}$ and $[-1, 7] \rm\ km\ s^{-1}$, respectively.  (d--f) HI integrated intensity map in the velocity range $[-48, -40] \rm\ km\ s^{-1}$, $[-15, -7] \rm\ km\ s^{-1}$ and $[-1, 7] \rm\ km\ s^{-1}$, respectively. }
    \label{fig:multiwavelength2}
\end{figure*}

Additional CO and HI integrated intensity maps are presented in Fig.~\ref{fig:multiwavelength2}.

Based on the MWISP CO data and Gaia DR3  \citep{gaia_mission, gaia_dr3_summary, gaia_edr3_am, gaia_ap}, we measured the distance to the potential bubble structure in the velocity range of $-15$ to $-7\, \mathrm{km\,s^{-1}}$, following the method described in \citet{distance_fit_method}.
To simplify the analysis, we manually selected three regions, labeled A, B, and C in Fig.~\ref{fig:multiwavelength2}(b), where Gaia stars were identified as the on-cloud stellar sample. These regions correspond to areas in the $\rm ^{13}CO$ integrated intensity map (-15 to $-7\, \mathrm{km\,s^{-1}}$) exceeding five times the noise level. Stars with $A_G$ errors smaller than 0.01 were excluded, as their small uncertainties would lead to disproportionately large weights. For the off-cloud sample, stars located in regions of the $\rm ^{12}CO$ integrated intensity map below 1 sigma were selected.
Using the off-cloud sample, we fitted a monotonically increasing extinction function to derive the baseline for $A_G$ extinction, accounting for the gradual extinction increase caused by interstellar diffuse material. This was achieved using the \texttt{IsotonicRegression}\footnote{\url{https://scikit-learn.org/dev/modules/generated/sklearn.isotonic.IsotonicRegression.html}} method from the Python package \texttt{scikit-learn}. After subtracting the baseline, the extinction in the on-cloud regions was modeled using a step function with five parameters: the cloud distance ($D$), the extinction $A_G$ ($\mu_1$) and standard deviation ($\sigma_1$) of foreground stars, and the extinction $A_G$ ($\mu_2$) and standard deviation ($\sigma_2$) of background stars.
We sampled the posterior distribution of these parameters using the MCMC method implemented in the \texttt{emcee}\footnote{\url{https://emcee.readthedocs.io/en/stable/}} package.

\begin{figure*}[p]
    \centering
    \includegraphics[width=0.9\textwidth]{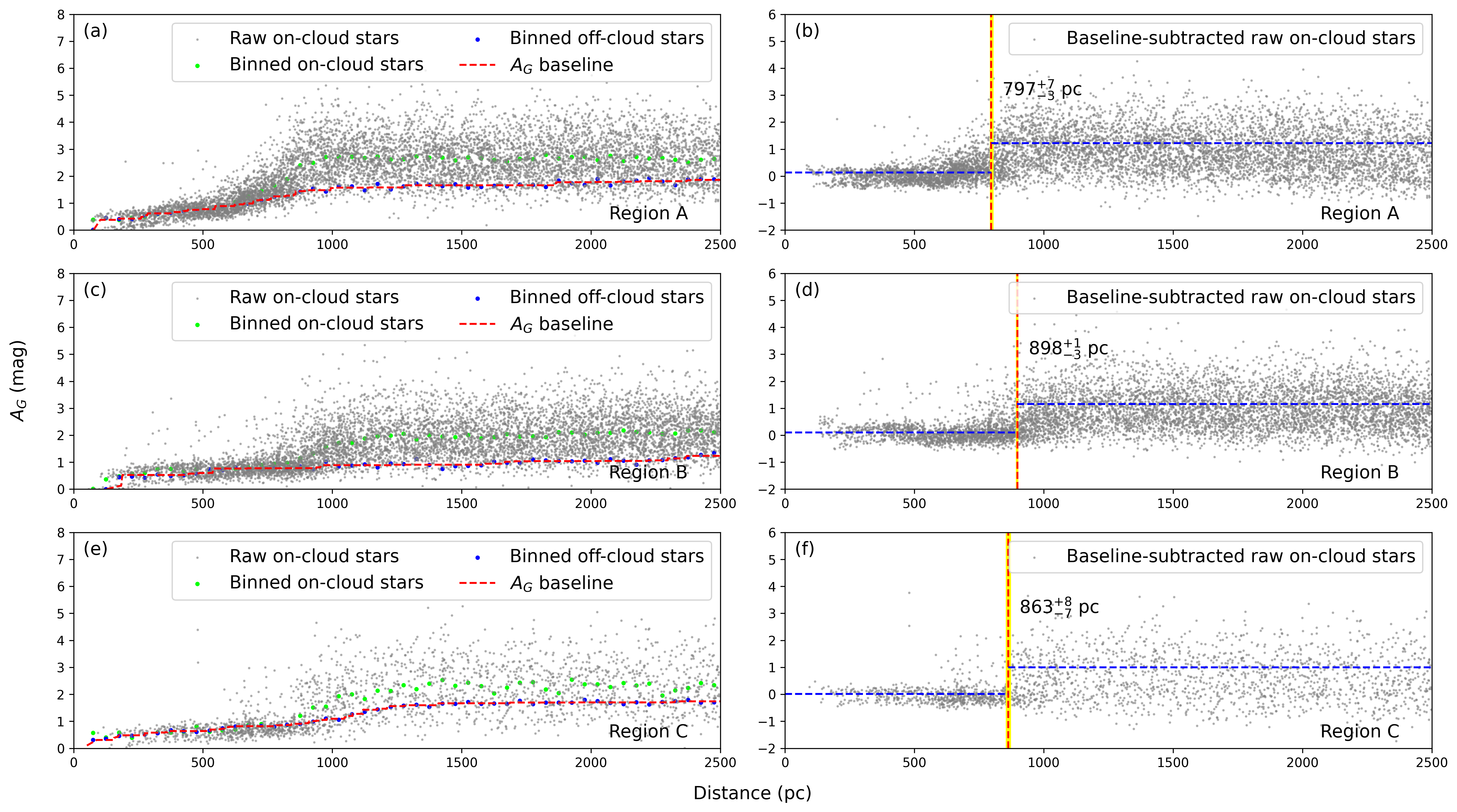}
    \caption{
        Distance estimation using Gaia extinction data. (a) Gray points show the raw on-cloud sample in Region A, while the red line shows the baseline fitted using off-cloud stars. Green and blue points indicate the binned on-cloud and off-cloud stars, respectively, used for visualization only (excluded from fitting). (b) Gray points represent the on-cloud sample in Region A after baseline subtraction. The red line and yellow band indicate the estimated distance and corresponding $1\,\sigma$ statistical uncertainty. Blue lines represent the average extinction of foreground and background stars. Panels (c–f) present similar plots as (a) and (b) for Regions B and C, respectively.
    }
    \label{fig:distance_fit}
\end{figure*}

In Fig.~\ref{fig:distance_fit}, we present the distance measurement results. The derived distances are  $797^{+7}_{-3} \, \mathrm{pc}$,  $898^{+1}_{-3} \, \mathrm{pc}$, and  $863^{+8}_{-7} \, \mathrm{pc}$  for Regions A, B, and C, respectively. The errors shown here represent only the statistical uncertainties of the model. As discussed in \citet{distance_fit_method}, the systematic error in the distances is approximately 5\%. For Region A, several velocity components with relatively strong intensities are observed at slightly smaller (in absolute value) velocities compared to the measured velocity component.  These components suggest the presence of foreground molecular clouds, which may obscure the target cloud and introduce additional uncertainties in the distance estimation.

\end{multicols}

\clearpage
Zhen Cao$^{1,2,3}$,
F. Aharonian$^{4,5}$,
Axikegu$^{6}$,
Y.X. Bai$^{1,3}$,
Y.W. Bao$^{7}$,
D. Bastieri$^{8}$,
X.J. Bi$^{1,2,3}$,
Y.J. Bi$^{1,3}$,
W. Bian$^{9}$,
A.V. Bukevich$^{10}$
% ...
(The LHAASO Collaboration)
$^{1}$ Key Laboratory of Particle Astrophysics \& Experimental Physics Division \& Computing Center, Institute of High Energy Physics, Chinese Academy of Sciences, 100049 Beijing, China\\
$^{2}$ University of Chinese Academy of Sciences, 100049 Beijing, China\\
$^{3}$ TIANFU Cosmic Ray Research Center, Chengdu, Sichuan,  China\\
$^{4}$ Dublin Institute for Advanced Studies, 31 Fitzwilliam Place, 2 Dublin, Ireland \\
$^{5}$ Max-Planck-Institut for Nuclear Physics, P.O. Box 103980, 69029  Heidelberg, Germany\\
$^{6}$ School of Physical Science and Technology \&  School of Information Science and Technology, Southwest Jiaotong University, 610031 Chengdu, Sichuan, China\\
$^{7}$ School of Astronomy and Space Science, Nanjing University, 210023 Nanjing, Jiangsu, China\\
$^{8}$ Center for Astrophysics, Guangzhou University, 510006 Guangzhou, Guangdong, China\\
$^{9}$ Tsung-Dao Lee Institute \& School of Physics and Astronomy, Shanghai Jiao Tong University, 200240 Shanghai, China\\
$^{10}$ Institute for Nuclear Research of Russian Academy of Sciences, 117312 Moscow, Russia\\
% ...

\end{document}